# The solutions of Nonlinear Evaluation equations via Hermite Approximation

*Zehra Pınar[a]   Turgut Öziş[b]*

[a]*Namık Kemal University, Faculty of Arts and Science, Department of Mathematics, 59030 Merkez-Tekirdağ, Turkey*

[b]*Ege University, Science Faculty, Department of Mathematics, 35100 Bornova-İzmir, Turkey*

Abstract: It is well recognized that new types of exact travelling wave solutions to nonlinear partial differential equations can be obtained by modifications of the methods which are in hand. In this study, we extend the class of auxiliary equations using Hermite differential equation so the solution space of nonlinear partial differential equations is extended too. The proposed Hermite differential equation plays an important role in quantum mechanics, probability theory, statistical mechanics, and in solutions of Laplace's equation in parabolic coordinates. Consequently, we introduce new exact travelling wave solutions of some physical systems in terms of the solutions of the Hermite differential equation.

Key Words: Hermite differential equation, BBM equation, Klein-Gordon equation, travelling wave solutions.

[a] Corresponding author: Zehra Pınar (zpinar@nku.edu.tr )

1. **Introduction**

The inspection of nonlinear wave phenomena of diffusion, convection, dispersion and dissipation appearing in engineering is of great interest from both mathematical and physical points of view. In most case, the theoretical modeling based on nonlinear partial differential equations (NLPDEs) can accurately describe the wave dynamics of many physical systems. Of critical importance is to find closed form solutions for NLPDEs of physical significance. This could be a very complicated task and, in fact, is not always possible since in various realistic problems in physical systems. So, searching for some exact physically significant solutions is an important topic because of wide applications of NLPDEs in biology, chemistry, physics, fluid dynamics and other areas of engineering.

Recently, to determine the solutions of nonlinear evolution equations, many exact solutions of various auxiliary equations have been utilized [4-10]. The most common properties of some of these methods are based on the polynomial approximations in $\infty$ - norm. But, the choice of norm can extensively manipulate the outcome of the problem. That is to say that the polynomial approximation to a given continuous function in a one norm need not carry any resemblance to the polynomial approximation in another norm. As a result, the choice of norm will be executed by the sense in which the given continuous function must be "good" approximation for many practical problems. Therefore "good" approximation in the 2-norm ($L^2$) is closely related to the concept of orthogonality and this, consequently leads to concept of inner product and may leads to the solutions of specific orthogonal polynomials which form complete orthogonal sets in $L^2$. In this manner, it is known that new applications of orthogonal polynomials from the higher-order polynomial classes and their connections to other areas of mathematics, physics, and engineering have emerged during the past few years. The von Neumann theory of self-adjoint extensions of Lagrangian formally symmetric differential expressions was applied by M.A. Naimark [11], N.I. Akhiezer [12], M.G. Krein, and I.M. Glazman [14], an important class of periodic spectral problems is that of self-adjoint operators and its most important applications are the spectral analysis of the classical second-order differential equations of Jacobi, Laguerre, and Hermite [15].

In this paper, we will examine the differences of the choice of different auxiliary equation which is Hermite equation for determining the solutions of the nonlinear evolution equation in consideration and more we search for additional forms of new exact solutions of nonlinear differential equations which satisfying Hermite equation.

## 2. Methodology of The Hermite Approximation

Let us have a nonlinear partial differential equation

$$P(u, u_x, u_t, u_{xx}, u_{xt}, u_{tt}, ...) = 0 \qquad (1)$$

and let by means of an appropriate transformation this equation be reduced to nonlinear ordinary differential equation

$$Q(u, u_\zeta, u_{\zeta\zeta}, u_{\zeta\zeta\zeta}, ...) = 0. \qquad (2)$$

For large class of the equations of the type (eq. (1)) have exact solutions which can be constructed via finite series

$$u(\zeta) = \sum_{i=0}^{N} a_i z^i(\zeta) \qquad (3)$$

Here, $a_i, (i = 0,1,...,N)$ are constants to be further determined, $N$ is an integer fixed by a balancing principle and elementary function $z(\zeta)$ is the solution of the Hermite differential equation. It is worth to point out that we happen to know the general solution(s), $z(\zeta)$, of the Hermite differential equation.

By definition of classical series solution; the series solution of Eq.(2) consists of substituting the series (3) in the differential equation and then attempting to determine what the coefficients $a_i, (i=0,1,...,N)$ must be in order that the series will satisfy the differential equation (2). But, if the equation (i.e. eq.(2)) is nonlinear, then determination of the coefficients $a_i, (i=0,1,...,N)$ not simple as in power series solutions due to elementary functions $z(\zeta)$ in series solution in eq.(3) and this lead us the following approximation:

*Approximation steps*:

A) Substitute Eq.(3) into ordinary differential equation (2) to determine the coefficients $a_i, (i=0,1,...,N)$ with the aid of symbolic computation.

B) Insert predetermined coefficients $a_i$ and elementary function $z(\zeta)$ of the Hermite differential equation into Eq.(3) to obtain travelling wave solutions of the nonlinear partial differential equation in consideration.

It is very apparent that if elementary function, $z(\xi)$, is orthogonal function which form complete orthogonal sets in $L^2$ then, the solution series will be convergent series therefore the series (7) will converge rapidly. In this study, we use the Hermite differential equation

$$\frac{d^2z}{d\zeta^2} = 2\zeta\left(\frac{dz}{d\zeta}\right) + \lambda z(\zeta) \qquad (4)$$

so in this way the new solutions will be obtained .

## 3. Application of the Hermite Approximation

### 3.1. Klein-Gordon Equation

Firstly, we consider the Klein-Gordon equation which produces the classical waves and also is used as a test problem by a number of researchers. The Klein-Gordon equation is a relativistic version of the Schrödinger equation and describes the spinless pion, a composite particle. The Klein–Gordon equation for a free particle has a simple plane wave solution. The generalized nonlinear Klein–Gordon equation is given by

$$\frac{\partial^2 u}{\partial t^2} - a^2 \frac{\partial^2 u}{\partial x^2} + \alpha u - \beta u^n = 0, n > 1 \qquad (5)$$

where $a, \alpha, \beta$ are real constant. When takes the values 2 and 3, Eq. (5) becomes the nonlinear Klein–Gordon equation with quadratic and cubic nonlinear term, respectively. To find the travelling wave solutions for Eq. (5), we use the wave variable $\zeta = \mu(x-ct)$, where $c \neq 0$ and $\mu \neq 0$. The wave variable $\xi$ carries Eq. (5) into the ordinary differential equation

$$\mu^2(c^2-a^2)u'' + \alpha u - \beta u^n = 0. \qquad (6)$$

From balancing principle and using (5), we have $N=2$, therefore the ansatz yields

$$u(\zeta) = g_0 + g_1 z(\zeta) + g_2 z^2(\zeta) \qquad (7)$$

where $z(\zeta)$ is a solution(s) of Eq. (4). Hence, substituting Eqs.(7) and (4) into Eq.(5) and letting each coefficient of $z(\zeta)$ to be zero, we obtain

$$-3\beta g_0 g_2^2 - 3\beta g_2 g_1^2 = 0,$$
$$-6\beta g_0 g_1 g_2 - \beta g_1^3 = 0,$$
$$2\mu^2 c^2 g_2 \lambda + \alpha g_2 - 2a^2\mu^2 g_2 \lambda - 3\beta g_2 g_0^2 - 3\beta g_0 g_1^2 = 0,$$
$$-4a^2\mu^2 g_2 \zeta \frac{dz(\zeta)}{d\zeta} + 4\mu^2 c^2 g_2 \zeta \frac{dz(\zeta)}{d\zeta} + \mu^2 c^2 g_1 \lambda - a^2\mu^2 g_1 \lambda + \alpha g_1 - 3\beta g_0^2 g_1 = 0,$$
$$2\mu^2 c^2 g_1 \zeta \frac{dz(\zeta)}{d\zeta} - 2a^2\mu^2 g_1 \zeta \frac{dz(\zeta)}{d\zeta} + 2\mu^2 c^2 g_2 \left(\frac{dz(\zeta)}{d\zeta}\right)^2 - 2a^2\mu^2 g_2 \left(\frac{dz(\zeta)}{d\zeta}\right)^2 - \beta g_0^3 + \alpha g_0 = 0$$

**Case 1:** In the event of $z'(\zeta) = h$, where $h$ is constant, the solution of system of equations is

$g_0 = 0, g_1 = 0, g_2 = g_2$ and in the case, the solution of Hermite equation is

$$z(\zeta) = \_C2 e^{\zeta\sqrt{\lambda}} + \_C1 e^{-\zeta\sqrt{\lambda}} - \frac{2h\zeta}{\lambda}$$

Substituting the above coefficients into ansatz (7) with the solution of Hermite equation, we obtain one of new solution of Klein-Gordon equation. The Figure 1 is given for the special values of parameters.

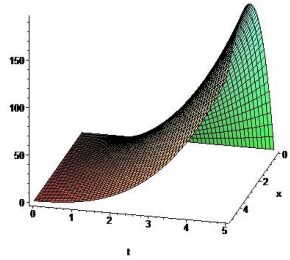

**Figure 1.** For $c=0.8, \mu=0.1, g_2=xt, h=2, \_C1=1, \_C2=2, \lambda=0.6$, graph of solution of Eq.(5)

**Case 2:** If it is considered that

$$z'(\zeta) = 2\_C1\left(KummerM\left(\frac{3}{2}+\frac{\lambda}{4},\frac{3}{2},\zeta^2\right) - KummerM\left(\frac{1}{2}+\frac{\lambda}{4},\frac{3}{2},\zeta^2\right)\right)\left(\frac{1}{2}+\frac{\lambda}{4}\right) + \_C1 KummerM\left(\frac{1}{2}+\frac{\lambda}{4},\frac{3}{2},\zeta^2\right)$$
$$2\_C2\left(\frac{\lambda}{4} KummerM\left(\frac{3}{2}+\frac{\lambda}{4},\frac{3}{2},\zeta^2\right) - KummerM\left(\frac{1}{2}+\frac{\lambda}{4},\frac{3}{2},\zeta^2\right)\right)\left(\frac{1}{2}+\frac{\lambda}{4}\right) + \_C2 KummerM\left(\frac{1}{2}+\frac{\lambda}{4},\frac{3}{2},\zeta^2\right)$$

the solutions of system of equations are

*i.*

ii.
$$\_C1 = \frac{-\frac{1}{4}\_C2\lambda\left(-4KummerU\left(\frac{1}{2}+\frac{\lambda}{4},\frac{3}{2},\zeta^2\right)+2KummerU\left(\frac{3}{2}+\frac{\lambda}{4},\frac{3}{2},\zeta^2\right)+\lambda KummerU\left(\frac{3}{2}+\frac{\lambda}{4},\frac{3}{2},\zeta^2\right)\right)}{-KummerU\left(\frac{1}{2}+\frac{\lambda}{4},\frac{3}{2},\zeta^2\right)+2KummerU\left(\frac{3}{2}+\frac{\lambda}{4},\frac{3}{2},\zeta^2\right)+\lambda KummerU\left(\frac{3}{2}+\frac{\lambda}{4},\frac{3}{2},\zeta^2\right)}$$

$$\alpha = -\frac{\mu^2\lambda(-c^2+a^2)}{2}, \beta = -\frac{\mu^2\lambda(-c^2+a^2)}{2g_0^2}, g_2 = \frac{3g_1^2}{2g_0}$$

iii.
$$\_C1 = \frac{-\frac{1}{4}\_C2\lambda\left(-4KummerU\left(\frac{1}{2}+\frac{\lambda}{4},\frac{3}{2},\zeta^2\right)+2KummerU\left(\frac{3}{2}+\frac{\lambda}{4},\frac{3}{2},\zeta^2\right)+\lambda KummerU\left(\frac{3}{2}+\frac{\lambda}{4},\frac{3}{2},\zeta^2\right)\right)}{-KummerU\left(\frac{1}{2}+\frac{\lambda}{4},\frac{3}{2},\zeta^2\right)+2KummerU\left(\frac{3}{2}+\frac{\lambda}{4},\frac{3}{2},\zeta^2\right)+\lambda KummerU\left(\frac{3}{2}+\frac{\lambda}{4},\frac{3}{2},\zeta^2\right)}$$

$$\alpha = -\mu^2\lambda(-c^2+a^2), \beta = -\frac{\mu^2\lambda(-c^2+a^2)}{g_0^2}, g_1 = 0$$

iv.
$$g_1 = g_2 \left(\begin{array}{l}-\_C1KummerM\left(\frac{3}{2}-\zeta^2,\frac{3}{2},\zeta^2\right)+2\zeta^2\_C1KummerM\left(\frac{3}{2}-\zeta^2,\frac{3}{2},\zeta^2\right)-2\zeta^2\_C1KummerM\left(\frac{1}{2}-\zeta^2,\frac{3}{2},\zeta^2\right)\\+\zeta^2\_C2KummerU\left(\frac{3}{2}-\zeta^2,\frac{3}{2},\zeta^2\right)-2\zeta^4\_C2KummerU\left(\frac{3}{2}-\zeta^2,\frac{3}{2},\zeta^2\right)-2\zeta^2\_C2KummerU\left(\frac{1}{2}-\zeta^2,\frac{3}{2},\zeta^2\right)\end{array}\right)\Big/\zeta$$

$$\alpha = -8\mu^2\zeta^2(-c^2+a^2), \lambda = -4\zeta^2, g_0 = 0$$

$$\_C1 = \frac{-\frac{1}{4}\_C2\lambda\left(-4KummerU\left(\frac{1}{2}+\frac{\lambda}{4},\frac{3}{2},\zeta^2\right)+2KummerU\left(\frac{3}{2}+\frac{\lambda}{4},\frac{3}{2},\zeta^2\right)+\lambda KummerU\left(\frac{3}{2}+\frac{\lambda}{4},\frac{3}{2},\zeta^2\right)\right)}{-KummerU\left(\frac{1}{2}+\frac{\lambda}{4},\frac{3}{2},\zeta^2\right)+2KummerU\left(\frac{3}{2}+\frac{\lambda}{4},\frac{3}{2},\zeta^2\right)+\lambda KummerU\left(\frac{3}{2}+\frac{\lambda}{4},\frac{3}{2},\zeta^2\right)}$$

$$\alpha = -\mu^2\lambda(-c^2+a^2), g_0 = 0, g_1 = 0$$

v.
$$g_0 = \frac{\sqrt{30}(-1+\sqrt{21})g_1}{80}\left(\begin{array}{l}-4\lambda\_C1KummerM\left(\frac{1}{2}+\frac{\lambda}{4},\frac{3}{2},\zeta^2\right)+2\lambda\_C2KummerU\left(\frac{3}{2}+\frac{\lambda}{4},\frac{3}{2},\zeta^2\right)\\+\lambda^2\_C2KummerU\left(\frac{3}{2}+\frac{\lambda}{4},\frac{3}{2},\zeta^2\right)+4\lambda\_C1KummerM\left(\frac{3}{2}+\frac{\lambda}{4},\frac{3}{2},\zeta^2\right)\\+8\_C1KummerM\left(\frac{3}{2}+\frac{\lambda}{4},\frac{3}{2},\zeta^2\right)-4\_C2\lambda KummerU\left(\frac{1}{2}+\frac{\lambda}{4},\frac{3}{2},\zeta^2\right)\end{array}\right)\Big/\sqrt{\lambda(-1+\sqrt{21})}$$

$$g_2 = -\frac{\sqrt{30}(-1+\sqrt{21})6\lambda g_1}{5}\Bigg/\left(\sqrt{\lambda(-1+\sqrt{21})}\left(\begin{array}{l}-\frac{6}{5}\lambda\_C1(-1+\sqrt{21})KummerM\left(\frac{1}{2}+\frac{\lambda}{4},\frac{3}{2},\zeta^2\right)+\frac{6}{5}\lambda\_C1(-1+\sqrt{21})KummerM\left(\frac{3}{2}+\frac{\lambda}{4},\frac{3}{2},\zeta^2\right)\\-12\lambda\_C1KummerM\left(\frac{1}{2}+\frac{\lambda}{4},\frac{3}{2},\zeta^2\right)+\frac{12}{5}(-1+\sqrt{21})\_C1KummerM\left(\frac{3}{2}+\frac{\lambda}{4},\frac{3}{2},\zeta^2\right)\\+12\lambda\_C1KummerM\left(\frac{3}{2}+\frac{\lambda}{4},\frac{3}{2},\zeta^2\right)+24\_C1KummerM\left(\frac{3}{2}+\frac{\lambda}{4},\frac{3}{2},\zeta^2\right)\\+\frac{3}{10}(-1+\sqrt{21})\lambda^2\_C2KummerU\left(\frac{3}{2}+\frac{\lambda}{4},\frac{3}{2},\zeta^2\right)-\frac{6}{5}\lambda\_C2(-1+\sqrt{21})KummerU\left(\frac{1}{2}+\frac{\lambda}{4},\frac{3}{2},\zeta^2\right)\\+3\_C2\lambda^2KummerU\left(\frac{3}{2}+\frac{\lambda}{4},\frac{3}{2},\zeta^2\right)+\frac{3}{5}(-1+\sqrt{21})\lambda\_C2KummerU\left(\frac{3}{2}+\frac{\lambda}{4},\frac{3}{2},\zeta^2\right)\\-12\_C2\lambda KummerU\left(\frac{1}{2}+\frac{\lambda}{4},\frac{3}{2},\zeta^2\right)+6\_C2\lambda KummerU\left(\frac{3}{2}+\frac{\lambda}{4},\frac{3}{2},\zeta^2\right)\end{array}\right)\right)$$

The figures of these solutions are given in Figure 2, respectively.

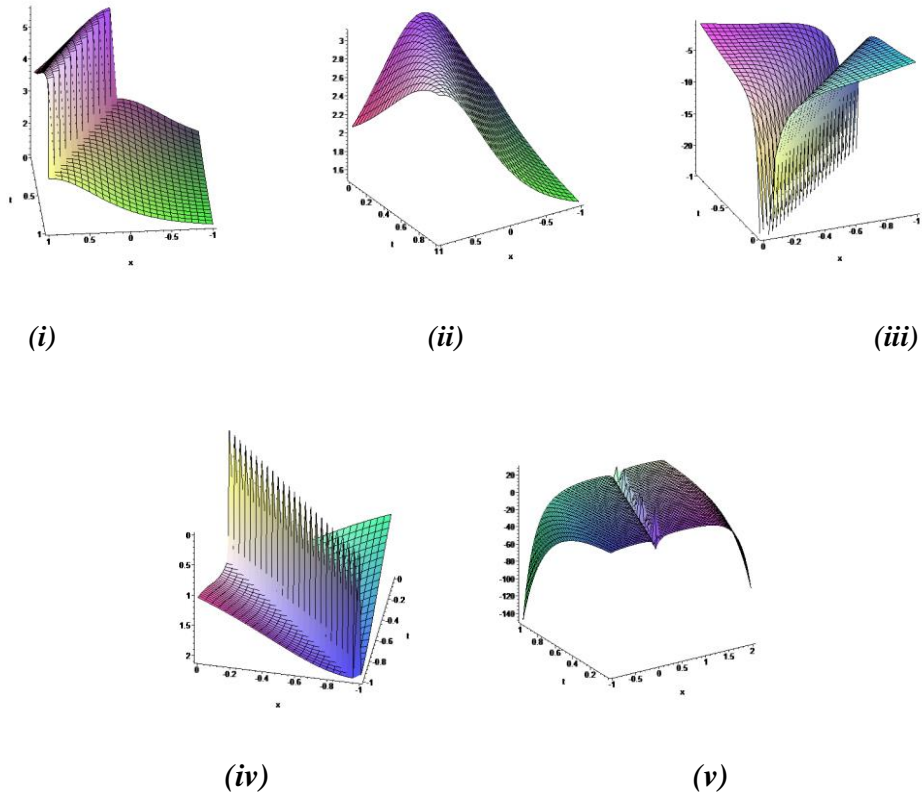

*(i)* *(ii)* *(iii)*

*(iv)* *(v)*

**Figure 2.** The solutions of Eq(5) are given respectively.

**Case 3:** If it is considered that $z'(\zeta) = A\sin(\zeta) + B\cos(\zeta)$, the solution of Hermite equation is

$$z(\zeta) = \_C2 e^{\zeta\sqrt{\lambda}} + \_C1 e^{-\zeta\sqrt{\lambda}} - \frac{(-2B(\lambda+1)\zeta - 4A)\cos(\zeta) - 2(A(\lambda+1)\zeta - 2B)\sin(\zeta)}{(\lambda+1)^2}$$

and the solutions of system of equations are as following;

**i.** $\alpha = \dfrac{\mu^2 c^2 \lambda}{2} - 50\mu^2 c^2 \zeta^2 + 50\mu^2 a^2 \zeta^2 - \dfrac{a^2 \mu^2 \lambda}{2}$, $g_0 = \dfrac{3g_2\left(2A\sin(\zeta)B\,RootOf(\_Z^2 + \sin(\zeta)^2 - 1) + A^2\sin(\zeta)^2 - B^2\sin(\zeta)^2 + B^2\right)}{50\zeta^2}$

$g_1 = \dfrac{g_2\left(A\sin(\zeta) + B\,RootOf(\_Z^2 + \sin(\zeta)^2 - 1)\right)}{5\zeta}$

$$\beta = \frac{1250}{9}\mu^2\zeta^4 \begin{pmatrix} (2c^2\lambda A\sin(\zeta)RootOf(\_Z^2+\sin(\zeta)^2-1))B - 2a^2\lambda A\sin(\zeta)RootOf(\_Z^2+\sin(\zeta)^2-1)B \\ -40c^2 AB\sin(\zeta)RootOf(\_Z^2+\sin(\zeta)^2-1)\zeta^2 + 40a^2 AB\sin(\zeta)RootOf(\_Z^2+\sin(\zeta)^2-1)\zeta^2 \\ +20a^2 A^2\sin(\zeta)^2\zeta^2 - 20c^2 B^2\zeta^2 + 20a^2 B^2\zeta^2 + c^2\lambda A^2\sin(\zeta)^2 + 20c^2\zeta^2 B^2\sin(\zeta)^2 \\ -a^2\lambda A^2\sin(\zeta)^2 - 20c^2 A^2\sin(\zeta)^2\zeta^2 - 20a^2 B^2\sin(\zeta)^2\zeta^2 - \lambda c^2 B^2\sin(\zeta)^2 + c^2\lambda B^2 \\ +\lambda a^2 B^2\sin(\zeta)^2 - a^2\lambda B^2 \end{pmatrix} \Bigg/ \left( g_2^2 \begin{pmatrix} 6RootOf(\_Z^2+\sin(\zeta)^2-1)\begin{pmatrix} A^5\sin(\zeta)^5 B - 20A^3\sin(\zeta)^5 B^3 + 20A^3\sin(\zeta)^3 B^3 + \\ 6A\sin(\zeta)^5 B^5 - 12A\sin(\zeta)^3 B^5 + 6A\sin(\zeta)B^5 \end{pmatrix} \\ +A^6\sin(\zeta)^6 - 15A^4\sin(\zeta)^6 B^2 + 15A^4\sin(\zeta)^4 B^2 + 15A^2\sin(\zeta)^6 B^4 \\ -30A^2\sin(\zeta)^4 B^4 + 15A^2\sin(\zeta)^2 B^4 - B^6\sin(\zeta)^6 + 3B^6\sin(\zeta)^4 \\ -3B^6\sin(\zeta)^2 + B^6 \end{pmatrix} \right)$$

**ii.** $A = -\dfrac{B\cos(\zeta)}{\sin(\zeta)}$, $\alpha = -\dfrac{\lambda\mu^2(-c^2+a^2)}{2}$, $\beta = -\dfrac{2\lambda\mu^2 g_2^2(-c^2+a^2)}{9g_1^4}$, $g_0 = \dfrac{3g_1^2}{2g_2}$

**iii.** $\alpha = 8\mu^2 c^2\zeta^2 - 8\mu^2 a^2\zeta^2$, $\lambda = -4\zeta^2$, $g_0 = 0$, $g_1 = -\dfrac{g_2\left(A\sin(\zeta) + RootOf(\_Z^2 + \sin(\zeta)^2 - 1)B\right)}{\zeta}$

*iv.* $A = -\dfrac{B\cos(\zeta)}{\sin(\zeta)}, \alpha = -2\lambda\mu^2(-c^2+a^2), g_0 = 0, g_1 = 0$

*v.* $A = -\dfrac{B\cos(\zeta)}{\sin(\zeta)}, \alpha = -\lambda\mu^2(-c^2+a^2), \beta = -\dfrac{\lambda\mu^2(-c^2+a^2)}{g_0^2}, g_1 = 0$

*vi.* $B = 0, \alpha = -\dfrac{\lambda\mu^2(-c^2+a^2)}{2}, \beta = -\dfrac{2\lambda\mu^2 g_2^2(-c^2+a^2)}{9g_1^4}, g_0 = \dfrac{3g_1^2}{2g_2}, \mu = -\dfrac{\pi}{-x+ct}$

*vii.* $B = 0, c = \dfrac{\alpha xt + \sqrt{2\alpha t^2 \lambda \pi^2 a^2 - 2\lambda \pi^2 \alpha x^2 + 4\lambda^2 \pi^4 a^2}}{\alpha t^2 + 2\lambda \pi^2}, g_0 = 0, g_1 = 0, \mu = -\dfrac{\pi}{-x + \left(\dfrac{\alpha xt + \sqrt{2\alpha t^2 \lambda \pi^2 a^2 - 2\lambda \pi^2 \alpha x^2 + 4\lambda^2 \pi^4 a^2}}{\alpha t^2 + 2\lambda \pi^2}\right)t}$

*viii.*
$B = 0, c = -\dfrac{\alpha xt + \sqrt{-\alpha t^2 \lambda \pi^2 a^2 + \lambda \pi^2 \alpha x^2 + \lambda^2 \pi^4 a^2}}{-\alpha t^2 + \lambda \pi^2}, g_1 = 0, \mu = -\dfrac{\pi}{-x - \left(\dfrac{\alpha xt + \sqrt{-\alpha t^2 \lambda \pi^2 a^2 + \lambda \pi^2 \alpha x^2 + \lambda^2 \pi^4 a^2}}{-\alpha t^2 + \lambda \pi^2}\right)t}$,

$\beta = -\dfrac{\lambda \pi^2 \left(-\dfrac{\alpha xt + \sqrt{-\alpha t^2 \lambda \pi^2 a^2 + \lambda \pi^2 \alpha x^2 + \lambda^2 \pi^4 a^2}}{-\alpha t^2 + \lambda \pi^2} + a^2\right)}{\left(-x - \left(\dfrac{\alpha xt + \sqrt{-\alpha t^2 \lambda \pi^2 a^2 + \lambda \pi^2 \alpha x^2 + \lambda^2 \pi^4 a^2}}{-\alpha t^2 + \lambda \pi^2}\right)t\right)^2 g_0^2}$

The solutions of Eq(5) are shown in Figure 3, respectively.

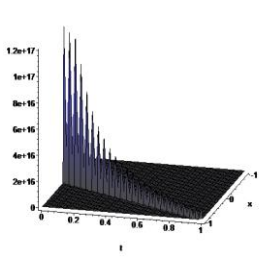
*(i)*

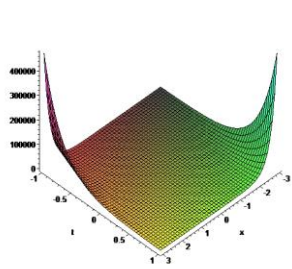
*(ii)*

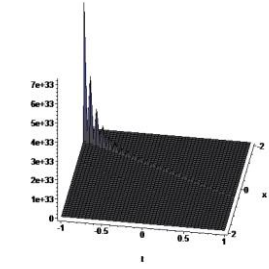
*(iii)*

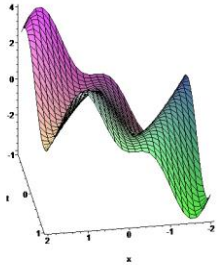
*(iv)*

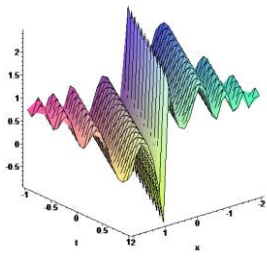
*(v)*

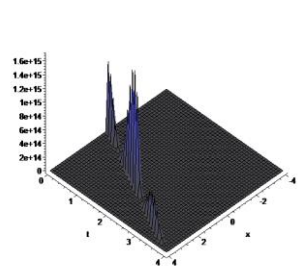
*(vi)*

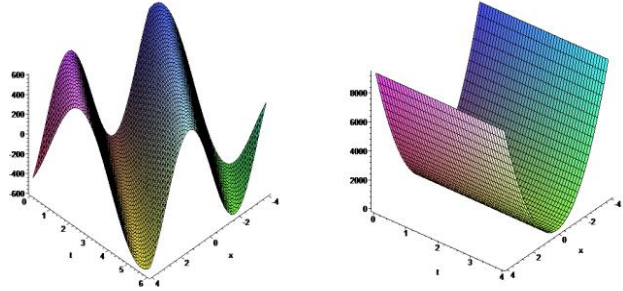

*(vii)* *(viii)*

**Figure 3.** Graph of solution of Eq.(5) for the subcases respectively.

### 3.2. Benjamin–Bona–Mahony Equation

The improvement of the Korteweg–de Vries equation (KdV equation) was proposed by Benjamin, Bona and Mahony (1972) . This equation is used for modeling long surface gravity waves of small amplitude –propagating uni-directionally in 1+1 dimensions.This contrasts with the KdV equation, which is unstable in its high wavenumber components.

Generalized BBM equation is

$$u_t + u_x + au^n u_x + u_{xxx} = 0, n \geq 1 \qquad (8)$$

with a constant parameter $a$. Here, we try to obtain the solutions of Eq(8) via Hermite approximation. From balancing principle and using (5), we have $N=1$, therefore the ansatz yields

$$u(\zeta) = g_0 + g_1 z(\zeta) \qquad (9)$$

where $z(\zeta)$ is a solution(s) of Eq. (4).Hence, substituting Eqs.(9) and (4) into Eq.(8) and letting each coefficient of $z^i(\xi)\sqrt{a_2 z^2(\xi) + a_6 z^6(\xi)}, (0 \leq i \leq 8)$ to be zero, we obtain

$$2a\mu \frac{dz(\zeta)}{d\zeta} g_0 g_1 + 2\mu^3 \zeta \lambda = 0,$$

$$\frac{dz(\zeta)}{d\zeta} c + \frac{dz(\zeta)}{d\zeta} \mu + a\mu \frac{dz(\zeta)}{d\zeta} g_0^2 + \mu^3 \lambda \frac{dz(\zeta)}{d\zeta} + 2\mu^3 \frac{dz(\zeta)}{d\zeta} + 4\mu^3 \zeta^2 \frac{dz(\zeta)}{d\zeta} = 0,$$

**Case 1:** In the event of $z'(\zeta) = h$, the solution of system of equations is

$$g_0 = \frac{\mu^2 \lambda \zeta}{h\sqrt{-a\mu(c+\mu+\mu^3\lambda+3\mu^3+4\mu^3\zeta^2)}}, g_1 = \frac{\sqrt{-a\mu(c+\mu+\mu^3\lambda+3\mu^3+4\mu^3\zeta^2)}}{a}$$

and in the case, the solution of Hermite equation is

$$z(\zeta) = \_C2 e^{\zeta\sqrt{\lambda}} + \_C1 e^{-\zeta\sqrt{\lambda}} - \frac{2h\zeta}{\lambda}$$

Substituting the above coefficients into ansatz (9) with the solution of Hermite equation, we obtain one of new solution of BBM equation. The Figure 4 is given for the special values of parameters.

**Case 2:** If it is considered that

$$z'(\zeta) = 2\_C1\left(KummerM\left(\frac{3}{2}+\frac{\lambda}{4},\frac{3}{2},\zeta^2\right) - KummerM\left(\frac{1}{2}+\frac{\lambda}{4},\frac{3}{2},\zeta^2\right)\right)\left(\frac{1}{2}+\frac{\lambda}{4}\right) + \_C1 KummerM\left(\frac{1}{2}+\frac{\lambda}{4},\frac{3}{2},\zeta^2\right)$$

$$2\_C2\left(\frac{\lambda}{4}KummerM\left(\frac{3}{2}+\frac{\lambda}{4},\frac{3}{2},\zeta^2\right) - KummerM\left(\frac{1}{2}+\frac{\lambda}{4},\frac{3}{2},\zeta^2\right)\right)\left(\frac{1}{2}+\frac{\lambda}{4}\right) + \_C2 KummerM\left(\frac{1}{2}+\frac{\lambda}{4},\frac{3}{2},\zeta^2\right)$$

the solution of system of equations is

$$\_C1 = \frac{-\frac{1}{4}\lambda\left(2KummerU\left(\frac{3}{2}+\frac{\lambda}{4},\frac{3}{2},\zeta^2\right)\_C2 ag_1 g_0 + \_C2 ag_1 g_0 \lambda KummerU\left(\frac{3}{2}+\frac{\lambda}{4},\frac{3}{2},\zeta^2\right) - 4\_C2 ag_1 g_0 KummerU\left(\frac{1}{2}+\frac{\lambda}{4},\frac{3}{2},\zeta^2\right) + 8\zeta\mu^2\right)}{\left(ag_1 g_0 \left(2KummerM\left(\frac{3}{2}+\frac{\lambda}{4},\frac{3}{2},\zeta^2\right) + KummerM\left(\frac{3}{2}+\frac{\lambda}{4},\frac{3}{2},\zeta^2\right)\lambda - \lambda KummerM\left(\frac{1}{2}+\frac{\lambda}{4},\frac{3}{2},\zeta^2\right)\right)\right)}$$

$$c = -\mu - a\mu g_0^2 - \lambda\mu^3 - 2\mu^3 - 4\mu^3\zeta^2$$

Substituting the above coefficients into ansatz (9) with the solution of Hermite equation, we obtain one of new solution of BBM equation. The Figure 4 is given for the special values of parameters.

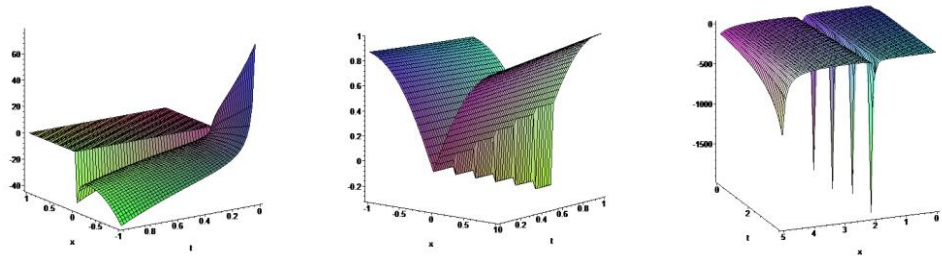

**Figure 4.** Graph of solution of Eq.(8) for each case of the BBM equation, respectively.

**Case 3:** If it is considered that $z'(\zeta) = A\sin(\zeta) + B\cos(\zeta)$, the solution of Hermite equation is

$$z(\zeta) = \_C2 e^{\zeta\sqrt{\lambda}} + \_C1 e^{-\zeta\sqrt{\lambda}} - \frac{(-2B(\lambda+1)\zeta - 4A)\cos(\zeta) - 2(A(\lambda+1)\zeta - 2B)\sin(\zeta)}{(\lambda+1)^2}$$

and the solution of system of equations is

$$c = \frac{\mu\left(-\zeta - ag_0^2\zeta + ag_0 g_1 A\sin(\zeta) + ag_0 g_1 B\cos(\zeta) - 2\mu^2\zeta - 4\mu^2\zeta^3\right)}{\zeta}, \lambda = -\frac{ag_0 g_1\left(A\sin(\zeta) + B\cos(\zeta)\right)}{\mu^2\zeta}$$

Substituting the above coefficients into ansatz (9) with the solution of Hermite equation, we obtain one of new solution of BBM equation. The Figure 4 is given for the special values of parameters.

Due to space limitation on the manuscript we only introduced two distinct examples to remark the applicability of Hermite equation regarding as new ansatz to obtain new travelling solutions of nonlinear physical systems.

**Conclusion**

As it is seen, the key idea of obtaining new travelling wave solutions for the nonlinear equations is using the exact solutions of different types equations as an ansatz. By means of Hermite equation with the wave transformation, Klein-Gordon equation has fourteen solutions in the total, while it has eight different solutions via extended auxiliary equation [9]. Therefore, using the solutions of Hermite type equation solution space of nonlinear partial differential equations is extended, so we have successfully obtained a number of new exact solutions of Klein-Gordon Equation and Benjamin–Bona–Mahony Equation by employing the solutions of the Hermit type equation regarding as an auxiliary equation in proposed method. Also, each case of the approximation, the nonlinear algebraic equation system has many solution sets. This property is good for engineers and physicians to simulate the solutions.

In this letter, we have obtained new solutions of the nonlinear equation in hand using the Hermite type equation for distinct cases. The presented method could lead to finding new exact travelling wave solutions for other nonlinear problems.